\documentclass{aa}
\usepackage{amssymb}
\usepackage{ulem}
\usepackage{color}
\input epsf.sty

\begin{document}

\title{The origin of the broad line region in active galactic nuclei}
\titlerunning{The origin of the BLR in AGN}

\author{Bo\.zena Czerny\inst{} \and Krzysztof Hryniewicz\inst{}}
\authorrunning{Czerny \& Hryniewicz}

\offprints{\email{krhr@camk.edu.pl}}

\institute{Nicolaus Copernicus Astronomical Center, Bartycka 18, 
	00-716 Warsaw, Poland}

\date{}

\abstract 
{}
{Although broad emission lines are the~most reliable signature of the nuclear activity of a galaxy and the location of the emitting material is well measured by the reverberation method, the physical cause of the formation of the broad line region remains unclear. We attempt to place some constraints on its origin.} 
{We study the properties of the accretion disk underlying the broad line region.} 
{We find that the effective temperature at the disk radius corresponding to the location of the broad line region, as inferred from the H$\beta$ line, is universal in all monitored sources and equal to 1000 K. This value is close to the limiting value that permits for the existence of the dust.} 
{The likely origin of the low ionization part of the broad line region is the strong local dusty wind from the disk. This wind becomes exposed to the irradiation by the central regions when moving higher above the disk surface and subsequently behaves like a failed wind, thus leading to a local mixture of inflow and outflow. This may provide the physical explanation of the turbulence needed both to  smooth the line profiles as well as provide additional mechanical heating.}

\maketitle

\section{Introduction}

Broad emission lines in the optical and UV band are the basic signature of active 
galactic nuclei and have been extensively studied for more than 50 years. These 
studies have allowed several key properties of the BLR to be identified: (i) the emitting medium is
located close to the accretion disk powering the AGN activity, and rarely
crosses the line of sight towards the nucleus; (ii) the medium is predominantly in 
Keplerian motion with additional velocity components (e.g., Done \& Krolik 1996;  
Kollatschny 2003; Collin et al. 2006);  (iii) the outer edge of the BLR
is delineated the dusty/molecular torus (Netzer \&  Laor 1993,
Suganuma et al. 2006); and (iv) the existence of 
BLR is likely to actually require the existence of the cold disk (e.g. Nicastro et al. 2003,
Czerny et al. 2004, Cao 2010). However, major uncertainties about
the geometry, dynamics, and physics of BLR still remain unknown (for a review, see e.g. Sulentic 
et al. 2000, Gaskell 2009). 

There is no such clear consensus on the mechanism determining the inner radius, 
or effective radius, of the BLR.
The monitoring of over 30 AGN has shown that, when the
contribution to the optical flux of the host galaxy is accounted for, there is a simple power-law relation between the BLR radius, as 
measured by the H$\beta$ line, and that the 
optical monochromatic luminosity, and the index of the power law is almost exactly equal 0.5, as expected
by simple ionization models.

The reason why this particular 
luminosity-dependent radius is chosen
by the sources remains unclear because reverberation studies have not yet returned information on the
3-D space density and velocity structure. There are several general suggestions of a wind type
outflow that either covers a large range of disk radii (Murray et al. 1995; 
Elitzur \& Shlosman 2006, 
Elitzur 2008), or mostly covers a narrow radial
range (Elvis 2000, Risaliti \& Elvis 2010), with a specific suggestion that the BLR forms at the radius where the gas pressure 
becomes comparable to the radiation--pressure. The wind outflow is expected to be either 
radiation pressure driven, or magnetically driven.

One aspect of the BLR scaling is, however, surprising. Among the reverberation-studied objects, 
there are both Seyfert 1 galaxies and narrow line Seyfert galaxies. The SEDs
of the two types of sources, when normalized at 5100 \AA~  flux, are different, including an 
expected difference in the ionizing flux (e.g. Wandel \& Boller 1998). Nevertheless, there seems to be no clear difference in the
BLR scalings of the two classes of objects, although the overall dispersion is large.
We attempt to explain this phenomenon in the present paper.  

\section{The BLR scaling law and the accretion disk}

The scaling of the BLR with luminosity discovered by Kaspi et al. (2000) has been studied by 
several authors (e.g. Peterson et al. 2004). Here we use
the result of Bentz et al. (2009), where the starlight contamination was carefully removed. 
The power-law index in the scaling between the BLR size and the monochromatic luminosity in 
their analysis is consistent with 0.5 within the errors. We thus fixed this slope at 0.5, 
and refitted their data again, obtaining the relation  
  
\begin{equation}
\log R_{BLR}[{\rm H}\beta] = 1.538 \pm 0.027 + 0.5 \log L_{44,5100},
\label{Bentz}
\end{equation}
where $R_{BLR}[{\rm H}\beta]$ is in light days and $L_{44,5100}$ is the monochromatic luminosity at 
5100 \AA~ measured
in units of $10^{44}$ erg s$^{-1}$. The value 0.027 is the error in the best-fit vertical 
normalization; the dispersion around the best fit is larger but still small at 0.21.

We assume that the disk is not strongly irradiated, and the outflow from the disk 
is not too strong, so the simple theory of the 
alpha-disk (Shakura \& Sunyaev 1973) applies. In this case, the monochromatic flux at 5100 \AA~ can be calculated
from the model, if the black hole mass and the accretion rate are known. Taking the numerical 
formula of Tripp et al. (1994) (corrected for a factor-of-two error, Nikolajuk, private communication),
we derive
\begin{equation}
\log L_{44,5100} = {2 \over 3} \log (M \dot M) - 43.8820 + \log \cos i,
\label{L5100}
\end{equation}
where the black hole mass, $M$, and the accretion rate, $\dot M$, are in g and g s$^{-1}$, respectively. 
This formula allows us to calculate the BLR radius from the Eq.~\ref{Bentz}, if the
product of ($M \dot M$) is known.

We now consider the effect of the inclination angle, $i$, on the measurements of the observed 
monochromatic flux, as well as the observed time delays. The estimate of the torus opening angle by Lawrence \& Elvis (2010) implies that the average value of the inclination angle of a broad line object is $i = 39.2^{\circ}$, and we use this same value for all objects. We also assume that the reflection 
comes predominantly from the part of the BLR behind the black hole, thus the measured time delay is related to the disk radius, $r$, underlying BLR 
using the formula
\begin{equation}
r = {R_{BLR}[{\rm H}\beta] \over 1 + \sin i}.
\label{eq:geom} 
\end{equation}

The Shakura-Sunyaev accretion disk theory allows us to calculate the effective temperature of the disk 
at that radius
\begin{equation}
T_{eff} = \left ({3GM \dot M \over 8 \pi r^3 \sigma_B}\right )^{0.25}
\label{Teff}
\end{equation}
where $\sigma_B$ is the Stefan-Boltzmann constant. Since BLR is far from the central 
black hole, the factor representing the inner boundary condition (including a 
spinning or non-spinning black hole) can be neglected.

When we combine Eqs.~\ref{Bentz}, \ref{L5100}, \ref{eq:geom}, and \ref{Teff}, the dependence on 
the unknown mass and accretion rate vanishes, and we obtain a single value for all the sources 
in the sample
\begin{equation}
T_{eff} = 995  \pm 74 {\rm K},
\end{equation}
where the error reflects the error in the constant in Eq.~\ref{Bentz}.

This value is interesting since it is close to the critical temperature at which the dust
can form. The distance is smaller than the distance to the dusty torus 
$R_{dust} \sim 0.4L_{45}^{1/2}$ pc (Nenkova et al. 2008; where $L_{45}$ stands for the bolometric 
luminosity in $10^{45}$ erg s$^{-1}$) since in our case the
temperature results only from the local disk dissipation instead of the irradiation by the 
central parts of AGN.

\section{Application to separate objects}

To assess the dispersion around this quantity in the Bentz et al. sample,  
we repeated the analysis using the individual measurements of
the time delay and the monochromatic starlight-subtracted  5100 
\AA~ luminosity provided by Bentz et al.
(2009) in their Table~8. We also separately studied a single source - NGC 5548 - since it 
had been monitored extensively over a number of years and 14 annual measurements for this source are available. 

We obtain the effective temperature of the disk related to BLR as before, for each 
measurement independently, converting the delay into
the BLR radius, and then disk radius (see Eq.~\ref{eq:geom}, as we assume the same 
geometrical factor for all sources) using Eqs.~\ref{L5100} and \ref{Teff}. 
The results are given in Table~\ref{tab:tempsey}.

The average value of the disk temperature obtained from the 48 measurements 
(all apart from NGC 5548) is
\begin{equation}
T_{eff} = 1030 \pm 61 {\rm K}
\end{equation}
and the dispersion in the measurements is 411 K for that sample.

Independently, we calculated the underlying disk temperature for the 14 measurements of 
NGC 5548 (see Table~\ref{tab:tempsey}). In this case, the average temperature was
\begin{equation}
T_{eff} = 956 \pm 56 {\rm K},
\end{equation}
with a dispersion 210 K. These values are fully consistent with the previous value inferred
directly from the scaling law.

We note that the dispersion in the measured value of the temperature is 
very low, only 0.16 in the logarithmic scale, comparable to the dispersion in the constant
in Eq.~\ref{Bentz}.

We checked whether the disk temperature underlying the BLR correlates with any AGN parameter. 
For that purpose,
we used the broad-band data fits for most of the monitored objects, provided by 
Vasudevan \& Fabian (2009). 
Since there was no one-to-one correspondence between several delay measurements and broad-band 
spectra measurements, in this case we used the mean delays measured 
for a given object from Bentz et al. (2009) and the mean values of the bolometric luminosity from Vasudevan \& Fabian (2009).
The results for the disk temperature did not differ from those obtained from all measurements 
($T_{eff} = 1029 \pm 88 $ K, with dispersion 441 K). 
As expected, there is no correlation between the disk effective temperature and any other parameter, e.g.,
 the black hole mass, the bolometric luminosity, and the Eddington ratio (the correlation 
coefficient was equal to -0.16, 0.08 and 0.26, respectively).   

We also calculated the ionizing flux for all of these sources, using the broad-band 
spectrum from Vasudevan \& Fabian (2009). For all the broad-band spectra, we calculated the 
ratio of the bolometric luminosity above 1 Ry
to the monochromatic luminosity at 5100 \AA. This dimensionless parameter accurately characterizes the expected
differences in the ionizing properties of continua in these sources. Our results are given in 
Table~\ref{tab:tempngc}. The average ratio is equal to 
$6.44 \pm 0.78$, but the dispersion is large (4.8). If this ratio is given in log space, 
the dispersion 
is equal to 0.40.

Therefore, the dispersion in the logarithm of the ionizing flux in those sources is much larger than
the dispersion in the scaling with monochromatic luminosity, 0.21, as well as the dispersion in 
the logarithm of the underlying disk temperature, 0.16. The ratio is also systematically higher for NLS1 galaxies ($10.6 \pm 2.6$) than for S1 objects ($5.5 \pm 0.7$). Interestingly, there seems to be a weak positive correlation between this ratio and $T_{eff}$.

\begin{table}
\caption{Effective temperature of the disk underlying BLR for sources from the reverberation sample.
Bolded values were computed from mean time delays from Bentz et al.}
\begin{minipage}[b]{.5\linewidth}
\centering
\begin{tabular}{cc}
\hline 
\hline
Name & $T_{eff}$ [K]\\
\hline
Mrk 335  &  1302 \\
  &  1731 \\
  &  {\bf 1427} \\
PG 0026+129  &  903 \\
PG 0052+251  &  913 \\
Fairall 9  &  1516 \\
Mrk 590  &  951 \\
  &  835 \\
  &  605 \\
  &  783 \\
  &  {\bf 749} \\
3C 120  &  962 \\
Ark 120  &  724 \\
  &  648 \\
  &  {\bf 808} \\
Mrk 79  &  1859 \\
  &  1318 \\
  &  1245 \\
  &  {\bf 1307} \\
PG 0804+761  &  690 \\
PG 0844+349a  &  1181 \\
Mrk 110  &  913 \\
  &  1109 \\
  &  633 \\
  &  {\bf 897} \\
PG 0953+414  &  858 \\
NGC 3227  &  1058 \\
  &  901 \\
  &  {\bf 787} \\
NGC 3516  &  995 \\
NGC 3783  &  1020 \\
NGC 4051  &  584 \\
NGC 4151  &  550 \\
PG 1211+143  &  826 \\
\hline
\end{tabular}
\end{minipage}
\begin{minipage}[b]{.47\linewidth}
\centering
\begin{tabular}{cc}
\hline 
\hline
Name & $T_{eff}$ [K]\\
\hline
PG 1226+032  &  986 \\
PG 1229+204  &  662 \\
NGC 4593  &  1890 \\
PG 1307+085  &  838 \\
IC 4329Ab  &  683$^*$ \\
Mrk 279  &  1227 \\
PG 1411+442  &  575 \\
PG 1426+015  &  748 \\
Mrk 817  &  1203 \\
  &  1274 \\
  &  707 \\
  &  {\bf 990} \\
PG 1613+658  &  1607 \\
PG 1617+175  &  756 \\
PG 1700+518  &  828 \\
3C 390.3  &  934 \\
Mrk 509  &  587 \\
PG 2130+099  &  1836 \\
NGC 7469  &  2408 \\ \hline
NGC 5548  &  828 \\
  &  694 \\
  &  940 \\
  &  968 \\
  &  1068 \\
  &  1094 \\
  &  864 \\
  &  987 \\
  &  792 \\
  &  788 \\
  &  772 \\
  &  1481 \\
  &  825 \\
  &  1289 \\
  &  {\bf 859} \\
 \hline
\end{tabular}
\end{minipage}
$^*$ time delay taken from the Winge et al. (1996).
\label{tab:tempsey}
\end{table}

\begin{table}
\caption{The dimensionless ratio of the luminosity above 1 Ry to the 
monochromatic luminosity at 5100 \AA. $LR$ is defined as 
$L_{1 {\rm Ry} - 100 {\rm keV}}/\lambda L_{\lambda}(5100{\rm \AA})$}
\begin{minipage}[b]{.45\linewidth}
\centering
\begin{tabular}{cc}
\hline 
\hline
Name & $LR$\\
\hline
3C 120  & 7.89 \\
3C 390.3  & 8.28 \\
Akn 120  & 2.92 \\
Fairall 9  & 2.34 \\
Mrk 110  & 13.33 \\
Mrk 279 (1)  & 8.71 \\
Mrk 279 (2)  & 9.10 \\
Mrk 279 (3)  & 8.73 \\
Mrk 335 (1)  & 13.89 \\
Mrk 335 (2)  & 23.14 \\
Mrk 509  & 3.17 \\
Mrk 590  & 6.75 \\
Mrk 79  & 5.12 \\
NGC 3227 (1)  & 6.28 \\
NGC 3227 (2)  & 4.13 \\
NGC 3516  & 2.10 \\
NGC 3783 (1)  & 4.97 \\
NGC 3783 (2)  & 4.05 \\
NGC 4051 (1)  & 6.49 \\
\hline
\end{tabular}
\end{minipage}
\begin{minipage}[b]{.45\linewidth}
\centering
\begin{tabular}{cc}
\hline 
\hline
Name & $LR$\\
\hline
NGC 4051 (2)  & 11.29 \\
NGC 4151 (1)  & 7.92 \\
NGC 4151 (2)  & 7.97 \\
NGC 4151 (3)  & 9.04 \\
NGC 4593  & 15.59 \\
NGC 5548  & 4.09 \\
NGC 7469  & 14.35 \\
PG 0052+251  & 2.91 \\
PG 0844+349  & 6.48 \\
PG 0953+414  & 4.07 \\
PG 1211+143 (1)  & 3.16 \\
PG 1211+143 (2)  & 2.56 \\
PG 1226+023  & 4.76 \\
PG 1229+204  & 4.34 \\
PG 1307+085  & 1.27 \\
PG 1411+442  & 0.37 \\
PG 1426+015  & 0.54 \\
PG 1613+658  & 1.92 \\
PG 2130+099  & 0.60 \\
\hline
\end{tabular}
\end{minipage}
\label{tab:tempngc}
\end{table}

\section{Discussion}

The observed scaling between the size of the BLR and the square root of 
the monochromatic flux at 5100 \AA~ measured according to the response of H$\beta$ to the continuum works
surprisingly well, although the ionizing flux is not expected to be proportional to
the monochromatic optical flux because of observed variety of SED shapes.

Using a simple analytic formula, we have demonstrated that the observed scaling corresponds to a fixed-temperature, non-irradiated Shakura-Sunyaev disk, independent of the black hole mass and
accretion rate. This temperature is about 1000 K, as derived
for several sources, and in agreement with the value obtained directly from the BLR 
scaling by Bentz et al.   
 
\begin{figure}
\epsfxsize=8.0cm
\epsfbox{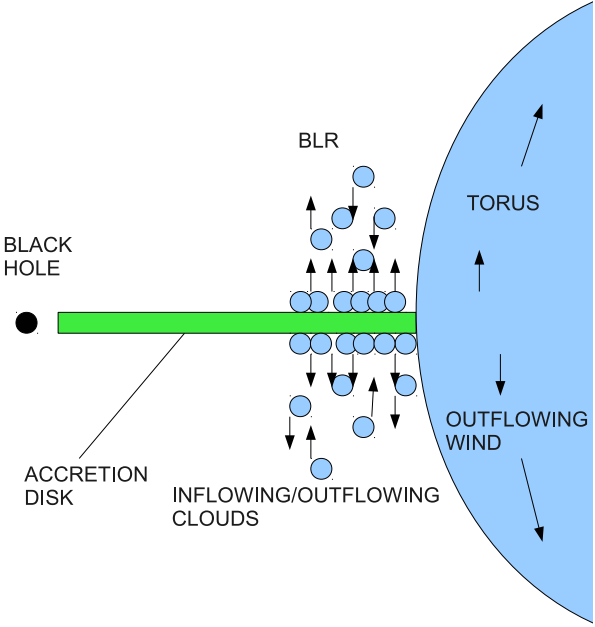}
\caption{The BLR region covers the range of the disk with an effective temperature lower than 
1000 K: the dusty wind rises and then breaks down when exposed to the radiation from the central source. 
The dusty torus is the disk range where the irradiation does not destroy the dust and the 
wind flows out.}
\label{fig:BLR}
\end{figure}

It is thus likely that the BLR forms when the local disk effective temperature becomes
low enough for dust to form in the disk atmosphere and the locally driven massive
outflow to develop, similar to the outflows observed from stars at the asymptotic 
giant branch. The pressure in the disk atmosphere, calculated with the code for the vertical disk structure (Rozanska et al. 1999), provides the conditions required for dust formation discussed by Elvis et al. (2002). However, this wind cannot achieve large heights, in contrast to stellar winds,
since at larger heights the wind is irradiated by the central source and the dust at these 
radial distances cannot exist, as this region is not yet at the distance of the dusty torus.
Therefore, the material, initially accelerated very efficiently, may partially fall back
owing to a decrease in the driving force. This would indicate that the BLR - or at least the 
low ionization line (LIL) part of it (Collin-Souffrin et al. 1988) - may consist of 
material, roughly in Keplerian motion, but with
clumps still outflowing and clumps flowing down. This chaotic motion will smear the expected
disk-like emission-line profile equally efficiently as in the assumption of a very extended flow, without
producing considerable asymmetry. We visualize the BLR geometry in Fig.~\ref{fig:BLR}.

This picture of a 'boiling' BLR region is consistent with two main lines of arguments, which have already 
been discussed in the literature. First, the observational arguments for both inflow and outflow in BLR were given in a number of papers
(e.g. Done \& Krolik 1996, Elvis 2000, Ferland et al. 2009, Gaskell 2009; Shapovalova et al. 2010).
Second, there has been considerable discussion of the heating efficiency of the LIL BLR: radiative heating
seems too low compared with the observed radiation field, and additional
mechanical heating seems to be required (Collin-Souffrin et al. 1988, Bottorff \& Ferland 2002, Shapovalova et al. 2004, 
Shapovalova et al. 2010).

Our scenario of local dust-driven wind with subsequent dust evaporation and inflow provides a physical basis for the development of this turbulent medium, with considerable velocity dispersion 
superimposed on the overall Keplerian motion of the underlying disk.   

This scenario can be quantitatively compared with alternative suggestions that the onset of the
BLR region is due to (i) the transition from the radiation pressure to gas pressure within the disk 
(Nicastro 2000) or (ii) the transition from non-self-gravitating to self-gravitating disk structure
(Collin \& Hure 2001). If the first option is the true underlying mechanism, we can estimate the position 
of this transition using the Shakura \& Sunyaev (1973) results. The transition takes place at the
radius $R_{ab}$  scaling as $M^{23/21} \dot m^{16/21}$ (in cm, not in gravitational radii), where 
$\dot m$ is the accretion rate in Eddington units. We can relate the black hole mass to the 
monochromatic luminosity again using the formula from Tripp et al. (1994). Thus, the expected
scaling of the BLR radius in this scenario can be represented as
\begin{equation}
R_{BLR}^{radpres} \propto L_{5100}^{23/28} \dot m ^{3/14},
\end{equation}   
such that the slope of the dependence on $L_{5100}$ is considerably steeper than 0.5, but not much steeper 
than for previous results inferred from the reverberation mapping. 

In the case of the self-gravity transition, we can estimate the transition radius
using the simple criterion that the local disk density should be equal to $GM/r
^3$. But since this region may occur in either region (b) or (c) (see Shakura \&
 Sunyaev 1973 for the definitions and analytic expression for the density), we have 
to calculate the expected trend for the two regions separately, obtaining
\begin{eqnarray}
R_{BLR}^{sg} \propto L_{5100}^{1/36} \dot m^{-83/54} ~~~({\rm region~~ b}), \nonumber \\
R_{BLR}^{sg} \propto L_{5100}^{-7/60} \dot m^{-37/90} ~~~({\rm region~~ c}),
\end{eqnarray}
so in both cases the dependence on the monochromatic luminosity differs strongly 
from our model. There is also an additional dependence on the accretion rate, which 
is weak (and positive) for the radiation pressure/gas pressure scenario, but 
much stronger for the self-gravity onset scenario. Monitoring results covering an even broader 
dynamical range of luminosity may be required to definitively rule out any of those alternative scenarios.

Within the framework of our BLR model, we can also estimate the radial size of the BLR, assuming that it covers the region 
from the radius where $T_{eff} = 1000 $K to the beginning of the dusty torus, 
$R_{dust} \sim 0.4L_{45}^{1/2}$ pc (Nenkova et al. 2008). The ratio of the two 
radii can be expressed in terms of the black hole mass, the accretion rate, and 
the accretion flow efficiency $\eta$ ($L_{bol} = \eta \dot M c^2$)
\begin{equation}
{r_{1000K} \over R_{dust}} = 0.03 {M_8^{1/6} \over \dot m^{1/6} \eta_{0.1}^{1/2}},
\end{equation}
where the black hole mass is expressed in $10^8 M_{\odot}$, and the efficiency in units of 0.1. 
Since the dependence on the black hole mass and the accretion rate is generally 
weak, in high accretion efficiency sources the outer radius is larger by 
a factor of 30 than the inner radius. However, if the accretion rate drops 
by a factor of $10^{-3}$, and, more importantly, the accretion efficiency 
drops by a factor of 100, then in our model the BLR is expected to reduce 
to a narrow ring ($r_{1000K} \approx  R_{dust}$). It is possible that this 
is the basic explanation of the relatively narrow BLR in several broad-line radio galaxies with clearly visible double peaks in the H$\beta$ 
line (see e.g. Lewis et al. 2010 and the references therein). 

If indeed the scenario outlined in the present paper is a correct interpretation, it means that
magnetic wind outflow is not the dominant mode of outflow from the disk. On the other hand, this type of outflow
is still likely to be present, perhaps being responsible for finally driving some material 
far away from the disk where high ionization lines form.

\begin{acknowledgements}
Part of this work was supported by grant 
NN 203 380136 of the Polish
State Committee for Scientific Research.
\end{acknowledgements}


\begin{thebibliography}{}

\bibitem[]{} Bottorff, M., Ferland, G., 2002, ApJ, 568, 581
\bibitem[]{} Bentz, M.C., Peterson, B.M., Netzer, H., Pogge, R.W., Vestergaard, M., 
       2009, ApJ, 705, 199
\bibitem[]{} Cao, X., 2010, arXiv1009.5043
\bibitem[]{} Collin, S., Kawaguchi, T., Peterson, B. M., Vestergaard, M., 2006, A\&A, 456, 75 
\bibitem[]{} Collin, S., Hure, J.-M., 2001, A\&A, 372, 50
\bibitem[]{} Collin-Souffrin, S., Dyson, J.E., McDowell, J.C.,  Perry, J.J., 1988, MNRAS, 232, 539
\bibitem[]{} Czerny, B.; Rozanska, A., Kuraszkiewicz, J., 2004, A\&A, 428, 39
\bibitem[]{} Done, C., Krolik, J.H., 1996, ApJ, 463, 144
\bibitem[]{} Elitzur, M. 2008, New Astron. Rev., 52, 274
\bibitem[]{} Elitzur, M., Shlosman, I., 2006, ApJ, 648, L101
\bibitem[]{} Elvis, M., 2000, ApJ, 545, 63
\bibitem[]{} Elvis, M., Marengo, M.,  Karovska, M., 2002, ApJ, 567, L107
\bibitem[]{} Ferland, G.J. et al., 2009, ApJ, 707, L82
\bibitem[]{} Gaskell, C.M., arXiv:0908.0386v2
\bibitem[]{} Gaskell, C.M., NewAR, 53, 140
\bibitem[]{} Kaspi, S., et al., 2000, ApJ, 533, 631
\bibitem[]{} Kollatschny, W.,  2003, A\&A, 407, 461
\bibitem[]{} Lawrence, A.,  Elvis, M., 2010, ApJ, 714, 561
\bibitem[]{} Lewis, K.T., Eracleous, M., Storchi-Bergmann, T., 2010, ApJS, 187, 416  
\bibitem[]{} Murray, N., Chiang, J., Grossman, S.A.,  Voit, G.M., 1995, ApJ, 451, 498
\bibitem[]{} Nenkova, M., Sirocky, M. M., Nikutta, R., Ivezic, Z., Elitzur, M. 2008, ApJ, 685, 160 
\bibitem[]{} Netzer, H., Laor, A., 1993, ApJ, 404, L51
\bibitem[]{} Nicastro, F., 2000, ApJ, 530, L65
\bibitem[]{} Nicastro, F., Martocchia, A., Matt, G.,2003, ApJ, 589, L13 
\bibitem[]{} Peterson, B.M., et al., 2004, ApJ, 613, 682
\bibitem[]{} Risaliti, G., Elvis, M.,  2010, A\&A, 516, A89
\bibitem[]{} Rozanska, A., Czerny, B., Zycki, P.T., Pojmanski, G., 1999, MNRAS, 305, 481
\bibitem[]{} Shakura, N.I., Sunyeav, R.A., 1973, A\&A, 24, 337
\bibitem[]{} Shapovalova, A.I., et al., 2004, A\&A, 422, 925
\bibitem[]{} Shapovalova, A.I., et al., 2010, A\&A, 509, 106
\bibitem[]{} Suganuma, M. et al. 2006, ApJ, 639, 46
\bibitem[]{} Sulentic, J. W., Marziani, P., Dultzin-Hacyan, D. 2000, ARA\&A, 38, 521
\bibitem[]{} Tripp, T.M., Bechold, J., Green, R.F., 1994, ApJ, 433, 533
\bibitem[]{} Vasudevan, R.V., Fabian, A.C., 2009, MNRAS, 392, 1124
\bibitem[]{} Winge, C., Peterson, B.M., Pastoriza, M.G., Storchi-Bergmann, T., 1996, ApJ, 469, 648 
\end{thebibliography}
\end{document}